\documentclass[10pt,conference]{IEEEtran}
\usepackage{tabularx}
\usepackage{booktabs}
\usepackage[labelfont=bf,format=plain,justification=raggedright,singlelinecheck=false]{caption}
\usepackage{xcolor}
\usepackage{comment}
\usepackage{subcaption,graphicx}
\usepackage{lastpage}
\usepackage{amsmath}

\usepackage{amsfonts}
\usepackage{amssymb}
\usepackage{enumerate}
\usepackage{mdframed}
\usepackage{array}
\usepackage{xcolor}
\usepackage{graphics}
\usepackage{tikz}
\usetikzlibrary{arrows}

\usepackage[noend]{algpseudocode}
\usepackage{algorithm}

\algnewcommand{\algorithmicgoto}{\textbf{go to}}%
\algnewcommand{\Goto}[1]{\algorithmicgoto~\ref{#1}}%

\usepackage{float}
  \usepackage[compact]{titlesec}
    \titlespacing{\section}{0pt}{0.5ex}{1ex}
    \titlespacing{\subsection}{0pt}{0.5ex}{0ex}
    \titlespacing{\subsubsection}{0pt}{0.5ex}{0ex}
\usepackage{amsmath}
\usepackage[justification=centering]{caption}
\usepackage[draft]{hyperref}

\IEEEoverridecommandlockouts 
\begin{document}
\newpage
\thispagestyle{empty}

\noindent\begin{minipage}{\textwidth}
    {\Huge\textbf{IEEE Copyright Notice}} \\ 

    \vspace{2cm}
    \Large{\copyright 2022 IEEE. Personal use of this material is permitted. Permission from IEEE must be obtained for all other uses, in any current or future media, including reprinting/republishing this material for advertising or promotional purposes, creating new collective works, for resale or redistribution to servers or lists, or reuse of any copyrighted component of this work in other works.}
\end{minipage}

\vspace{2cm} 

\noindent\begin{minipage}{\textwidth}
    
    \LARGE{Accepted for publication in: \textbf{IEEE Global Communications Conference (Globecom)} \\ \\
    DOI: \textbf{10.1109/GLOBECOM48099.2022.10000936}}
    
\end{minipage}

\newpage
\bstctlcite{IEEEexample:BSTcontrol}
\title{\textcolor{red}{\small{This paper has been accepted for publication in GLOBECOM 2022-2022 IEEE Global Communications Conference, pp. 2116-2121.}} \\ [2ex] CaMP-INC: Components-aware Microservices Placement for In-Network Computing Cloud-Edge Continuum \\
\thanks{This work was fully supported by CHIST-ERA program under the "Smart Distribution of Computing in Dynamic Networks (SDCDN)" 2018 call.}

\author{
    \IEEEauthorblockN{Soukaina Ouledsidi Ali\IEEEauthorrefmark{1},
                      Halima Elbiaze\IEEEauthorrefmark{1},
                      Roch Glitho\IEEEauthorrefmark{2},
		              Wessam Ajib\IEEEauthorrefmark{1}}
\IEEEauthorrefmark{1} Université du Québec à Montréal,
\IEEEauthorrefmark{2} Concordia University, Montreal, Canada
\\
ouledsidi\_ali.soukaina@courrier.uqam.ca, $\{$elbiaze.halima, ajib.wessam$\}@$uqam.ca, glitho@ece.concordia.ca } 
}

\maketitle
\begin{abstract}
Microservices are a promising technology for future networks, and many research efforts have been devoted to optimally placing microservices in cloud data centers. However, microservices deployment in edge and in-network devices is more expensive than the cloud. Additionally, several works do not consider the main requirements of microservice architecture, such as service registry, failure detection, and each microservice's specific database. This paper investigates the problem of placing components (i.e. microservices and their corresponding databases) while considering physical nodes' failure and the distance to service registries. We propose a Components-aware Microservices Placement for In-Network Computing Cloud-Edge Continuum (CaMP-INC). We formulate an Integer Linear Programming (ILP) problem with the objective of cost minimization. Due to the problem's $\mathcal{NP}$-hardness, we propose a heuristic solution. Numerical results demonstrate that our proposed solution CaMP-INC reduces the total cost by $15.8\%$ on average and has a superior performance in terms of latency minimization compared to benchmarks.
\end{abstract}


\IEEEpeerreviewmaketitle

\section{Introduction}
In the context of the rapidly growing number of internet- connected devices with a variety of distributed applications, including machine learning, the centralized cloud computing model has its limitations in meeting the emerging application requirements \cite{KULATUNGA2017106}. Cloud computing was originally designed for monolithic applications deployed in data centers with highly available resources. Also, contemporary multi-access edge computing aims to meet these new performance requirements by distributing application functionalities to edge devices, instead of running them solely in remote data centers. Mean- while, industry and academia have been paying close attention to In-Network Computing (INC) paradigm that leverages net- work devices for computing tasks. INC has been proven ef- fective in reducing latency and network traffic, and increasing throughput \cite{Dan2019}. Therefore, the computing infrastructure of the future spans the cloud, the core network and the edge. The heterogeneity of the devices in cloud-edge continuum with INC calls for the microservice architecture since microservices can be deployed in differently sized hosts \cite{8585089}. The success of microservice architecture entails five main requirements: (i) microservices replication, (ii) service registration, (iii) a dedicated database, (iv) resiliency and (v) load balancing.

First, to ensure microservices replication, a mechanism should be set to allow microservice scaling on different nodes based on incoming requests, \cite{hawilo2019exploring} and \cite{sampaio2019improving}. Second, to allow service discovery, the service registry plays a pivotal role in microservice architecture, \cite{hawilo2019exploring} and \cite{pallewatta2019microservices}. It is a database with the location of available microservices. When a new microservice is deployed, it must first register to this service registry. Third, each microservice should have its dedicated database in order to be decoupled from other microservices \cite{messina2016database}. Yet, they can be deployed in separate nodes since they can scale independently. Fourth, the distributed nature of microservices makes them more fault tolerant and resilient \cite{kakivaya2018service}. However, dependent microservices can be disturbed by physical nodes failure \cite{hawilo2019exploring}. To reinforce the resiliency, microservice architecture should be provided with a failure detection mechanism and recovery. Fifth, a load balancer is necessary to balance requests load among the same microservices \cite{pallewatta2019microservices} \cite{liu2019e3}.

The literature includes a few research works about microservices placement. The authors in~\cite{ding2022kubernetes} studied the dynamic competition and availability of microservices and the problem of shared dependency libraries among microservice instances. However, they only considered load balancing. The authors of \cite{kaur2022latency} introduced a novel microservice placement strategy considering the internal service composition in terms of communication between microservices. Paper  \cite{mpcsm} investigated the end-to-end (E2E) service latency estimation methods and how they enhance the microservices placement. Nonetheless, \cite{kaur2022latency}, and \cite{mpcsm} mainly focus on satisfying performance requirements and fails to address the microservice architecture requirements.
Existing microservices placement methods typically focus on them as a single component and ignore their databases and the service registry. Furthermore, many studies are restricted to either cloud \cite{ding2022kubernetes}, edge or INC paradigm and ignore the possibility of combining them. 

Considering the limitations of existing methods, this paper aims to ensure the QoS of delay constrained applications while minimizing the cost and taking into account microservice architecture requirements (i.e. namely requirements (ii), (iii), and (iv) previously mentioned) in In-Network Computing Cloud-Edge Continuum, and to the best of our knowledge, this is the first research work that addresses these limitations.


The remainder of this paper is organized as follows. Section~\ref{sysmod} introduces the system model, then Section~\ref{prob} presents the problem statement. In section~\ref{proposedmthd} we detail our proposed solution for microservices placement. Thereafter, in Section~\ref{evaluation}, we compare the performance of our solutions with benchmarks. Finally, Section~\ref{conclusion} concludes the paper.

\section{System Model}\label{sysmod}
\begin{figure*}[h]
\centering
\includegraphics[scale=0.3]{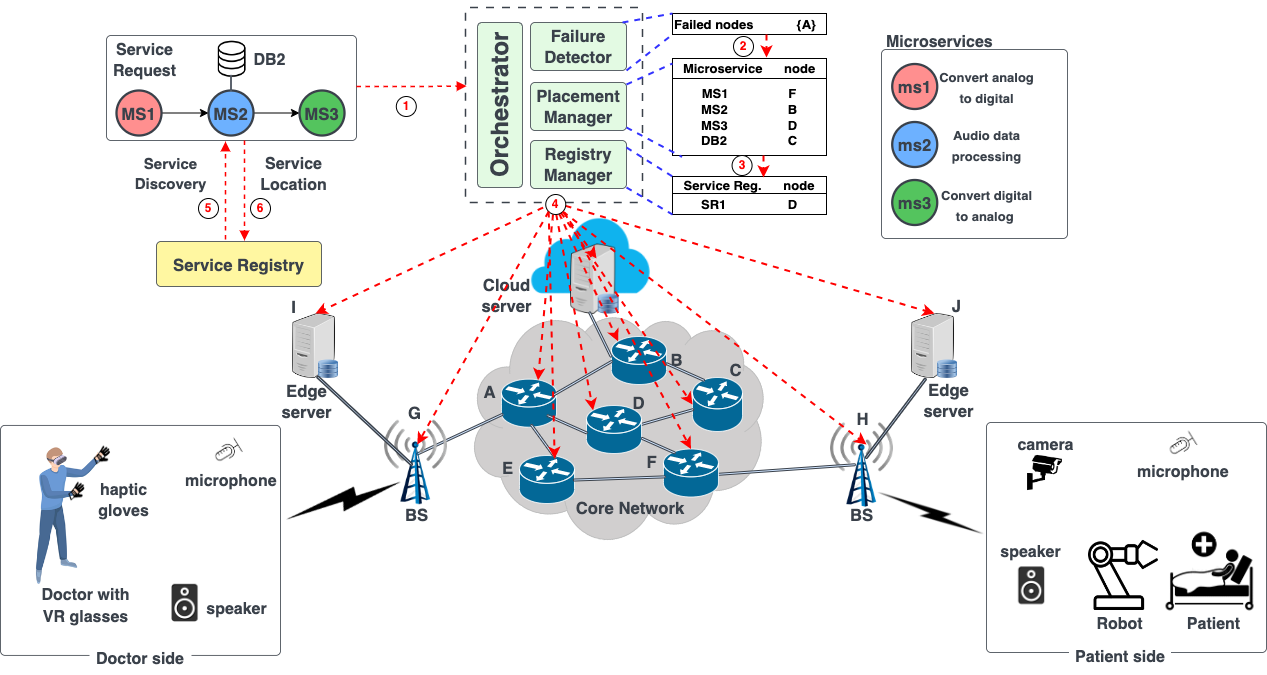}
    \caption{System model}
\label{system_model}
\end{figure*}

This paper studies the microservices placement problem in the cloud-edge continuum with INC to take advantage of the low access latency of edge and network nodes and the high computational power of the cloud. Our design goals include:
\begin{itemize}
    \item Each microservice should have a database component if it uses data as input or stores results in a database. 
    \item Resiliency mechanism to detect failed nodes and avoid using them for placement.
    \item Ensuring communication between microservices and service registry without violating delay threshold.
\end{itemize}

To better illustrate the system model, we consider the holopatient use case where doctors can explore the body, see visual indications, and verify health problems through remote examinations while going through the holographic patient. In addition, the holographic technology allows doctors to get haptic feedback from the patient side and also hear, see, and evaluate different pathologies and their symptoms in holographic patients. To have a pleasant experience for both doctor and patient, it is mandatory to choose the optimal placement of the processing to provide real-time interactions.

The content (i.e. in holopatient use case: microphone recordings, camera recordings, robot sensor data and haptic glove sensors data) is generated by content generator devices such as cameras, microphones, robots and haptic gloves. The end users send requests to run services on content. The service is a sequence of microservices that should be executed. Each microservice communicates with a private database through read/write operations. 
The microservices are deployed in servers and network devices. As shown in Fig.~\ref{system_model}, a service request arrives at the orchestrator (step 1), which is the component responsible for managing the microservices. There are three main elements in the orchestrator : (1) Failure Detector (FD), (2) Placement Manager (PM), and (3) Registry Manager (RM). FD periodically checks for detecting failures in the network. It periodically checks the network status and
keeps track of failed nodes in a list. Each time this list is updated, FD notifies the PM (step2). The PM then decides the placement of microservices and their databases on network nodes accordingly. After placing the chain of microservices, the RM verifies the distance between nodes hosting the new microsevices and the existing service registries (step 3). If the distance exceeds a certain threshold, then the RM decides to place a new service registry in the neighborhood of the microservices.
The orchestrator then distributes the microservices and the databases on the nodes decided by PM (step 4).
After placement, the service registry triggers newly deployed microservices for service discovery (step 5). Finally, the microservices send their location (i.e. IP address) to the service registry (step 6).

\section{Problem Statement }\label{prob}
Our problem is defined as dynamic placement of microservices and their dedicated databases while minimizing the total cost and satisfying the QoS requirements.

Let $V$ be the set of system devices and $U$ be the set of end users.  The physical network is represented by a graph $G = (N,E)$, where $N = V \cup U$ is the set of physical nodes and $E$ is the set of communication edges between nodes. Each device $k \in V$ has an amount of resources $r_k$.

\begin{table}[h]
    
    \begin{tabular}{|p{1.3cm}|p{6.4cm}|}
     \hline
     
     Notation & Description \\ [1ex] 
     \hline\hline 
     $V$ & set of all system devices  \\ [0.5ex] 
     $U$ & set of end users \\[0.5ex] 
     $E$ & set of all communication edges between devices  \\ [0.5ex]  
     $S$ & set of available microservices  \\ [0.5ex] 
     $DB$ & set of available databases  \\ [0.5ex]  
     $Q$ & set of services requested by end users\\  [0.5ex]  
     $Q_u$ & service request of end user $u \in U$ \\ [0.5ex]
     $N^u$ & set of microservices for request $Q_u \in Q$  \\ [0.5ex] 
     $DB^u$ & set of databases for request $Q_u \in Q$ \\ [0.5ex]  
     $E^u$ & set of  virtual edges for request $Q_u \in Q$  \\ [0.5ex] 
     $Z^u$ & set of devices hosting content for request $Q_u \in Q$  \\ [0.5ex] 
     $r_k$ & available resources in device $k \in V$  \\ [0.5ex] 
     $c_v$ & required resources for microservice $v \in S$  \\ [0.5ex] 
     $\phi_v$ & required resources for container packaging micro. $v \in S$  \\ [0.5ex]
     $v_f^u$ & the first microservice for request $Q_u \in Q$ \\ [0.5ex]
     $v_l^u$ & the last microservice for request $Q_u \in Q$ \\ [0.5ex]
     $\lambda_v$ &  cost of running an instance of microservice $v \in S$  \\ [0.5ex] 
     $\lambda_{db}$ &  cost of running an instance of database $db \in DB$  \\ [0.5ex]
     $\gamma_v$ &  deployment cost of microservice $v \in S$  \\ [0.5ex] 
     $\gamma_{db}$ &  deployment cost of database $db \in DB$  \\ [0.5ex] 
     $\eta_{v,k}$ & time to create a container for micro. $v$ in device $k \in V$ \\ [0.5ex]
     $\psi_{v,k}$ & time to release a container for micro. $v$ in device $k \in V$ \\ [0.5ex] 
     $\beta_{(p,q)}$ &  cost of using a physical link $(p,q) \in E$  \\ [0.5ex] 
     $d_{(p,q)}$ &  delay of physical link $(p,q) \in E$  \\ [0.5ex]
     $p_v$ &  processing delay of microservice $v \in S$ \\ [0.5ex] 
     $\Delta_{th}$ &  delay threshold \\ [0.5ex]
     $Reg_{v}$ &  delay between the service registry and the node hosting microservice $v \in S$ \\ [0.5ex]
     \hline
    \end{tabular}
    \caption{Table of notations}
    \label{table:1}
\end{table}

Let $S$ be the set of available microservices. Each microservice $v \in S$ might have a database $db \in DB$ and requires an amount of resources denoted $c_v$.
We assume that each microservice $v$ is packaged in a container that requires $\phi_v$ amount of resources.
The end user $u \in U$ service request is denoted $Q_u \in Q$ and modeled as a directed graph $Q_u = ( N^u, E^u)$, where $N^u$ represents the set of all microservices and their databases of $Q_u$ that should be placed on physical nodes $N$, and $E^u$ is the set of virtual edges. We consider two types of edges: (1) Edges between microservice (i.e. microservices order), (2) and Edges between each microservice and its dedicated database. 
We also define the following variables:
\begin{itemize}
    \item $v_f^u$: first microservice to be executed for request $Q_u$,
    \item $v_l^u$: last microservice to be executed for request $Q_u$,
    \item $Z^u$: the set of devices hosting content requested by $Q_u$.
\end{itemize}

Our objective is to minimize the total cost including operational cost, deployment cost and communication cost. Let $\lambda_{v}$ denotes the cost of running an instance of microservice $v$ and $\lambda_{db}$ the cost of running an instance of the database $db$. The operational cost is defined as :
\begin{equation}
    C_O = \sum\limits_{k \in N, \ v \in N^u} \lambda_{v} x_{v,k} + \sum\limits_{k \in N, \ db \in DB^u} \lambda_{db} y_{db, k}
\end{equation}
where $x_{v,k}$ = 1 if microservice $v$ is placed in device $k$ and $0$ otherwise and $y_{db,k}$ = 1 if database $db$ is placed in device $k$.

Deploying a new microservice $v$ entails a deployment cost $\gamma_{v}$ (i.e. license cost) per instance and deploying a new database $db$ costs $\gamma_{db}$. The total deployment cost is then: 
\begin{equation}
    C_D = \sum\limits_{k \in N, \ v \in N^u} \gamma_{v} x_{v,k} + \sum\limits_{k \in N, \ db \in DB^u} \gamma_{db} y_{db,k}
\end{equation}
We also consider $\beta_{(p,q)}$ as the cost of using a physical link $(p,q) \in E$ with $p,q \in N$ as the pair of devices hosting the microservices to be executed for the end user request $Q_u$. The communication cost includes also the link cost $\beta_{(z,k)}$ between device $z \in Z^u$ hosting the content requested by end user $Q_u$ and device $k \in V$ hosting the first microservice to be executed for $Q_u$ and the link cost $\beta_{(k,u)}$ between the device $k \in V$ hosting the last microservice to be executed for $Q_u$ and end user $u \in U$. Let $\beta_{(i,j)}$ denotes the link cost between the device hosting a microservice and a device hosting its database. Then the overall communication cost is:
\begin{eqnarray}
   \nonumber C_C &= \sum\limits_{\substack{(u,v) \in E^u \\ (p,q) \in E}} \beta_{(p,q)} m_{(p,q)}^{(u,v)}
    + \sum\limits_{\substack{(z,k) \in E \\ z \in Z^u}} \beta_{(z,k)} x_{v_f^u,k} 
    \\ &+  \sum\limits_{\substack{(k,u) \in E \\ u \in U}} \beta_{(k,u)} x_{v_l^u,k}
    + \sum\limits_{\substack{(v,db) \in E^u \\ (i,j) \in E}} \beta_{(i,j)} m_{(i,j)}^{(v,db)}
\end{eqnarray}
where $m_{(p,q)}^{(u,v)}$ = 1 if edge $(p,q) \in E$ hosts virtual link $(u,v) \in E^u$ and 0 otherwise.



The objective is to find the placement matrices: $\mathbf{X}= [x_{v,k}]$ of size $N^u \times V$, $\mathbf{Y}= [y_{db,k}]$ of size $DB^u \times V$ and $\mathbf{M}= [m_{(p,q)}^{(u,v)}]$ of size $E^u \times E$ that minimizes the total cost. Hence, the problem can be formulated as:
\begin{eqnarray}
&\underset{\mathbf{X,Y,M}}{\min} & C_{total} = C_O + C_D + C_C 
\label{problem}
\\
\nonumber
& \text{s. t.}
&\sum_{v \in N^u} (c_v + \phi_v) x_{v,k} + \\ & &
\sum_{db \in DB^u} (c_{db} + \phi_{db}) y_{db,k} \le r_k,
\forall k \in N
\label{C1}
\\ & &
\nonumber
\sum\limits_{\substack{(u,v) \in E^u \\ (p,q) \in E}} d_{(p,q)} m_{(p,q)}^{(u,v)} + 
\sum_{v \in N^u} p_v x_{v,k} + 
\\& &
\nonumber \sum_{(z,k) \in E} (p_{v_f^u} + d_{(z,k)}) x_{v_f^u,k} + 
\\ & &
\nonumber 
\sum_{(k,u) \in E} (p_{v_l^u} + d_{(k,u)}) x_{v_l^u,k} + 
\sum_{v \in N^u} (\eta_{v,k}+\psi_{v,k})
\\ & & 
+
\sum_{v \in N^u} Reg_v x_{v,k} \le \Delta_{th}  
, \forall k \in N
\label{C2}
\\ & &
\sum_{(u,v) \in E^u}  m_{(p,q)}^{(u,v)} = 1, \forall (p,q) \in E  
\label{C3}
\\ & &
\sum_{v \in N^u}  x_{v,k} = 1 , \forall k \in N
\label{C4}
\end{eqnarray}


Constraint \eqref{C1} guarantees that required resources by microservice $v \in N^u$, database $db \in DB^u$, the container packaging microservice $v$ and the container packaging the database $db$ do not exceed the capacity of the devices where they are located. Constraint \eqref{C2} guarantees that the response time to end user's service request does not exceed the predefined delay threshold $\Delta_{th}$. The delay constraint is the sum of the following terms: (i) the link delay of each physical link between each pair of devices hosting microservices and/or databases used to provide the requested service, (ii) the processing delay of these microservices, (iii) the communication delay between the content generator (i.e. the selected one to provide content to end user request) and the device hosting the first microservice and the processing delay of this microservice, (iv) the communication delay between the end user and the device hosting the last microservice and the processing delay of the latter (v) the time to create a container and release it $\eta_{v,k}$ and $\psi_{v,k}$ respectively for microservice $v$ on device $k$, and (vi) the delay between the service registry and the microservices. Constraints \eqref{C3} and \eqref{C4} are mapping constraints.


To prove the $\mathcal{NP}$-hardness of problem~\eqref{problem}, we reduce the known generalized assignment problem (GAP) \cite{cattrysse1992survey} to it.
Given $n$ bins and $m$ items, bin $i$ has a capacity $C_i$, $1$$\le$$i$$\leq$$n$, item $j$ has cost $c_{ij}$ with weight $w_j$ if assigned to bin $i$, $1$$\le$$j$$\leq$$m$. The GAP aims to minimize the total cost through assigning items to bins, subject to the bin capacities.
We consider a special case of problem~\eqref{problem} where only microservices and their databases are placed and there is no dependencies between them. Therefore, if the microservice or database $j$ is placed at device $i$, it consumes $w_j$ of resources and incurs a total cost of $c_{ij}$. To this end, problem~\eqref{problem} aims to minimize the total cost of placing microservices and databases, subject to the capacity $C_i$ of each microservice or database $i$. This special case of our problem is equivalent to the GAP. Hence, problem~\eqref{problem} is $\mathcal{NP}$-hard, due to the $\mathcal{NP}$-hardness of the GAP.

\section{Proposed method}\label{proposedmthd}
The proposed CaMP-INC solution consists of three main elements, namely (i) the placement manager (PM), which places microservices and their databases, (ii) the failure detector (FD) and (iii) the registry manager (RM), which decides if there is a need to place a new service registry.
\begin{algorithm}
\caption{Optimal Placement}\label{alg:OP}
\begin{algorithmic}[1]
\small
                \Require $G$: network graph, $q$ : EU request, $\Delta_{th}$: delay threshold
                \Ensure PlacementStrategy, PathWithMinCost
                    \State $ F \gets \Call{FailedNodes}{G}$
                    \State $ L \gets \Call{K-ShortestPaths}{G \setminus F, q.EU, q.CN, \Delta_{th} , K}$
                    \State $ MinCost \gets \infty $
                    \State $ C \gets \infty $
                    \For{$l \in L$}
                        \State $ P  \gets \emptyset $
                        \State $ BadPath \gets false $
                        \For{$ m \in q.S $}
                            \State $ lRank \gets \Call{rank}{l } $
                            \State $ BestNode \gets \Call{NodeWithHighestRank}{lRank } $
                            \If{$ BestNode = null $}
                                \State $ BadPath \gets true $
                                \State $ break $
                            \EndIf
                            \State $ P.add((m,BestNode)) $
                            
                        \EndFor
                        \If{$ BadPath = false $}
                            \State $ C \gets \Call{PlacementCost}{P } $
                            \If{$ C < MinCost $}
                                \State $ MinCost \gets C $
                                \State $ PlacementMin \gets P $
                                \State $ PathMin \gets l $
                            \EndIf
                        \EndIf

                    \EndFor
                    
                    \If{$ C = \infty $}
                        \State \Return $ \emptyset$
                    \Else
                        \If{$ \Call{RegistryDecision}{G \setminus F, PathMin, \Delta_{th}} $}
                            \For{$ n \in \Call{Neighbors}{PlacementMin}$}
                                \If{$ \Call{SatisfyDelay}{n, PlacementMin, \Delta_{th}} $}
                                    \State $ PlacementMin.add((Registry,n)) $
                                    \State $ break $
                                \EndIf
                                
                            \EndFor
                        \EndIf
                        \State \Return $ PlacementMin, PathMin $
                    \EndIf
        \end{algorithmic}
\end{algorithm}    
\subsection{Placement Manager (PM)}
The PM communicates with the FD and RM to decide on the placement of microservices, their databases, and, if necessary, a new service registry. The placement algorithm (Algorithm~\ref{alg:OP}) takes as input the network graph $G$, the end user request $q$ and the delay threshold $\Delta_{th}$. It starts by retrieving the failed nodes detected by the FD and then calculates $K$ shortest paths between the content node (i.e. q.CN) and the end user (i.e. q.EU) considering the delay threshold (lines $1$--$2$). For each path, node ranking \cite{page1999pagerank} is used to assign each microservice or its database (i.e. the components of microservices and databases chain q.S) to the node with the highest rank (lines $5$--$14$). If all microservices are assigned to nodes in the path, the placement cost is calculated (lines $15$--$16$). If the latter is minimal, then it is saved with the corresponding placement and path (lines $17$--$20$). An empty set is returned if no placement is found (lines $21$--$22$). Otherwise, the algorithm checks the need to place a new service registry using the RM. If it satisfies the delay threshold for each neighbour, then the registry is placed there. Finally, the placement with minimum cost and the path used in this placement are returned (lines $24$--$29$).

\subsection{Failure Detector (FD)}
The FD periodically checks the failed nodes as described in Algorithm~\ref{alg:FD}. It first sends a healthcheck message to each node in $G$ (line $2$). Then, if no acknowledgment is received, it marks the node as failed for the period and inform the PM by sending a set of failed nodes (lines $3$--$5$). The algorithm waits for a time period $t$ before rechecking (line $6$).

\begin{algorithm}
\caption{Failure Detection}\label{alg:FD}
\begin{algorithmic}[1]
\small
        \Require $G$: network graph, $t$: time period
        \For{$ node \in G $} \label{marker}
            \State $ \Call{HealthCheck}{node} $
            \If{$\neg \Call{Ack}{node} $}
                \State $ \Call{MarkFailed}{node} $
                \State inform placement manager
            \EndIf
        \EndFor
        \State $ \Call{Wait}{t} $
        \State \Goto{marker}
\end{algorithmic}
\end{algorithm}

\subsection{Registry Manager (RM)}
The RM uses Algorithm~\ref{alg:RD} to verify if there is a need for placing a new service registry. The input is the graph $G$, the path $path$ and the delay threshold $\Delta_{th}$. First, RM retrieves the locations of existing service registries (line $2$). Then, it calculates the delay between each node in the path and service registry, verifies the delay and returns the result (lines $4$--$10$).

\begin{algorithm}
\caption{Registry Decision}\label{alg:RD}
\begin{algorithmic}[1]
\small
    \Function{RegistryDecision}{$G$: network graph, $path$ : path, $\Delta_{th}$: delay threshold} : boolean
        \State $ L_{reg} \gets \Call{FindRegistries}{G} $
        \State $ exceededDelay \gets false $
        \For{$ reg \in L_{reg} $}
            \State $ D \gets \Call{Delay}{path,reg} $
            \If{$ D > \Delta_{th} $}
                \State $ exceededDelay \gets true $
            \EndIf
        \EndFor
        \If{$ exceededDelay = true$}
            \State \Return true
        \EndIf
        \State \Return false
    \EndFunction
\end{algorithmic}
\end{algorithm}

\begin{figure*}[t]
    \centering
  \begin{subfigure}{.5\linewidth}
    \centering
    \includegraphics[scale=0.195]{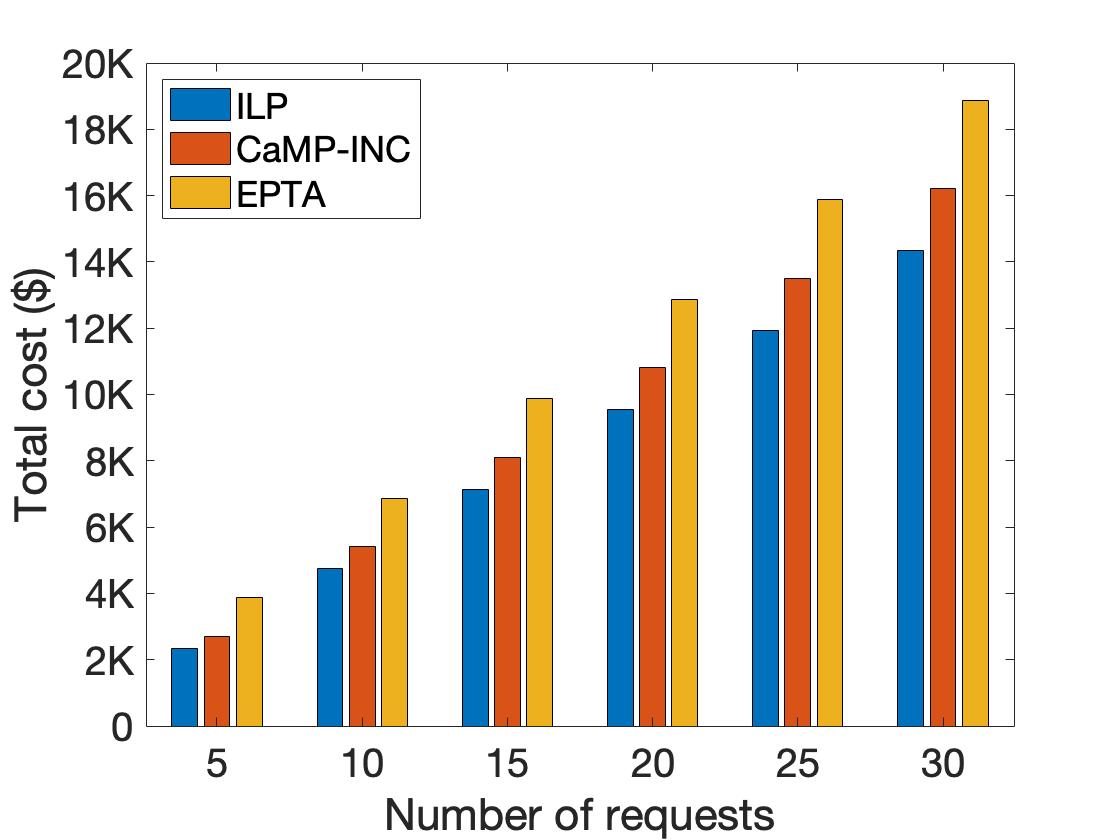}
    \caption{Total cost of placement}
    \label{total_cost}
  \end{subfigure}%
  \begin{subfigure}{.5\linewidth}
    \centering
    \includegraphics[scale=0.195]{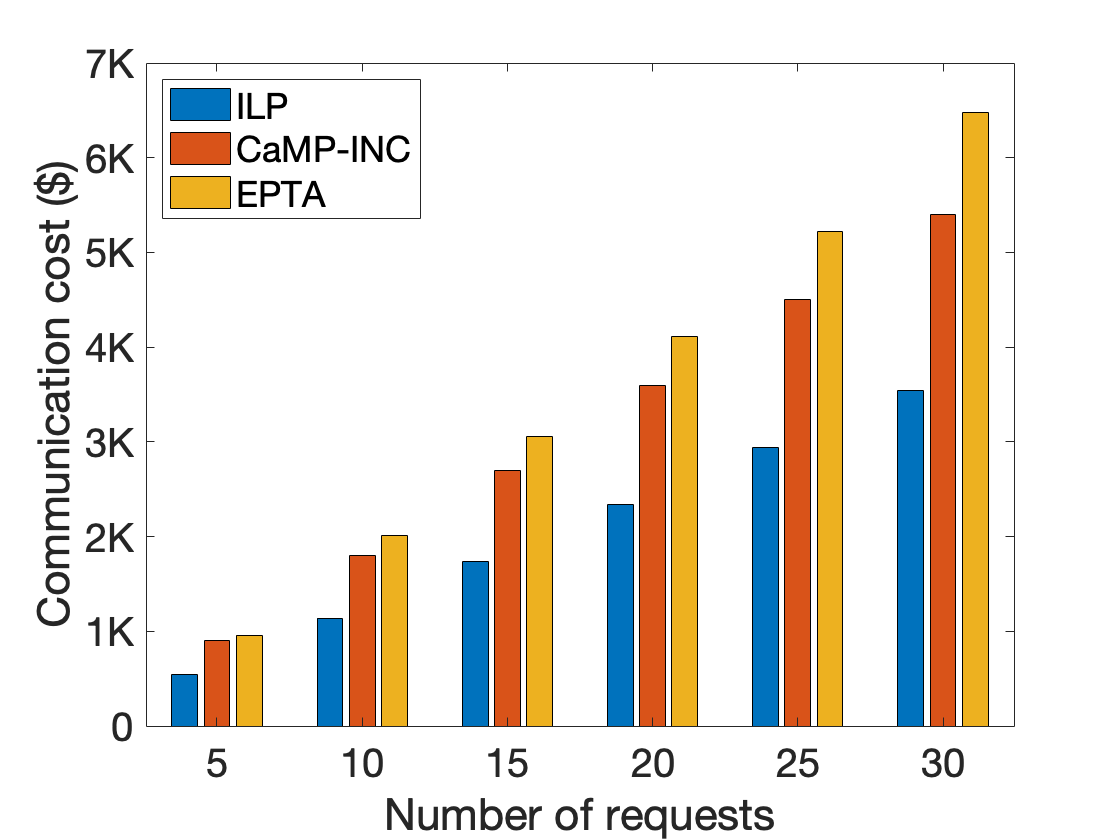
}
    \caption{Communication cost of placement}
    \label{communication_cost}
  \end{subfigure}
  \hspace{0em}
  \caption{Cost minimization}
\end{figure*}

\section{Evaluation}\label{evaluation}
In this section, we present our simulation and the performance comparison. The optimal solution (named ILP) is obtained using Gurobipy v9.5.1, and the heuristic algorithms are coded in Python. 
We simulated the network topology in figure \ref{system_model} with 8 users (i.e. 4 are content requesters and 4 are content generators), 2 base stations connecting the users to 2 edge servers and the core network consisting of 6 network devices, and a cloud server connected to the core network.
The physical nodes capacity varies in \{250, 500, 1000, 10000\} CPU cycles/s depending on if the physical node is respectively a network device, a base station, an edge server or a cloud server. The license cost of a microservice is $100\$$ and the number of microservices per request varies in $[3-5]$. The delay threshold varies in $[10-100]$ ms.
We variate the number of requests in \{5, 10, 15, 20, 25, 30\} and calculate the performance of the ILP, CaMP-INC and the state-of-art solution named Edge container Placement and Task assignment Algorithm (EPTA) \cite{EPTA}.


Fig. \ref{total_cost} depicts the total cost of different schemes for each number of requests. We observe that ILP can reduce the total cost by over 11\% and 24\% compared with CaMP-INC and EPTA, respectively. Furthermore, CaMP-INC represents a good trade-off between the ILP and EPTA. On the other hand, EPTA has the worst performance, as it prefers to select the nodes near to the controller, even if the link cost between the requester and the content is very high.
It can be observed in Fig. \ref{communication_cost} that the ILP outperforms CaMP-INC and EPTA, where EPTA has the highest communication cost. It is because microservices in EPTA communicate with each other through the controller, whereas in CaMP-INC, microservices can communicate directly with each other.

\begin{figure}
\centering
\includegraphics[scale=0.195]{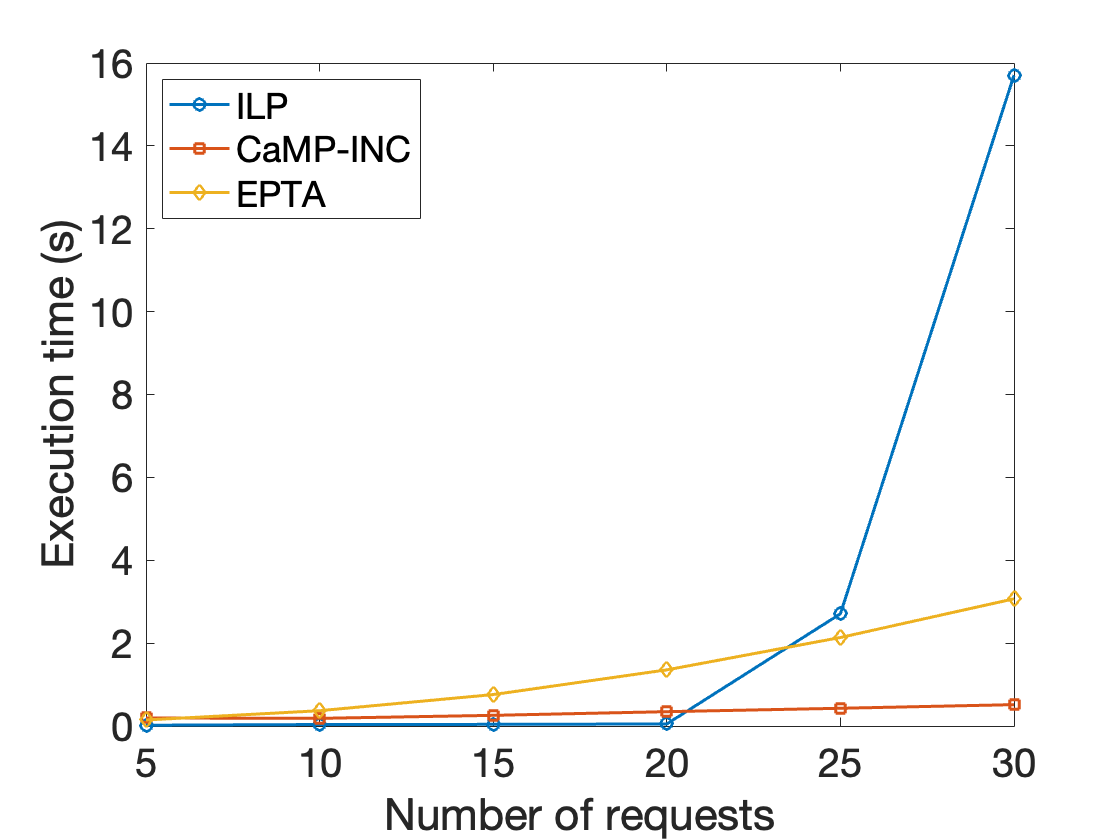}
    \caption{Execution time}
\label{execution_time}
\end{figure}

Fig. \ref{execution_time} shows that the execution time of ILP increases significantly with the increase in the number of requests. Starting from 35 requests, the execution time takes hours, making it impractical. Nevertheless, both heuristics have an acceptable execution time (i.e. under 4s). CaMP-INC outperforms EPTA as EPTA uses a light version of linear programming in the heuristic, which explains the behaviour of EPTA execution time.

\begin{figure*}
    \centering
  \begin{subfigure}{.5\linewidth}
    \centering
    \includegraphics[scale=0.195]{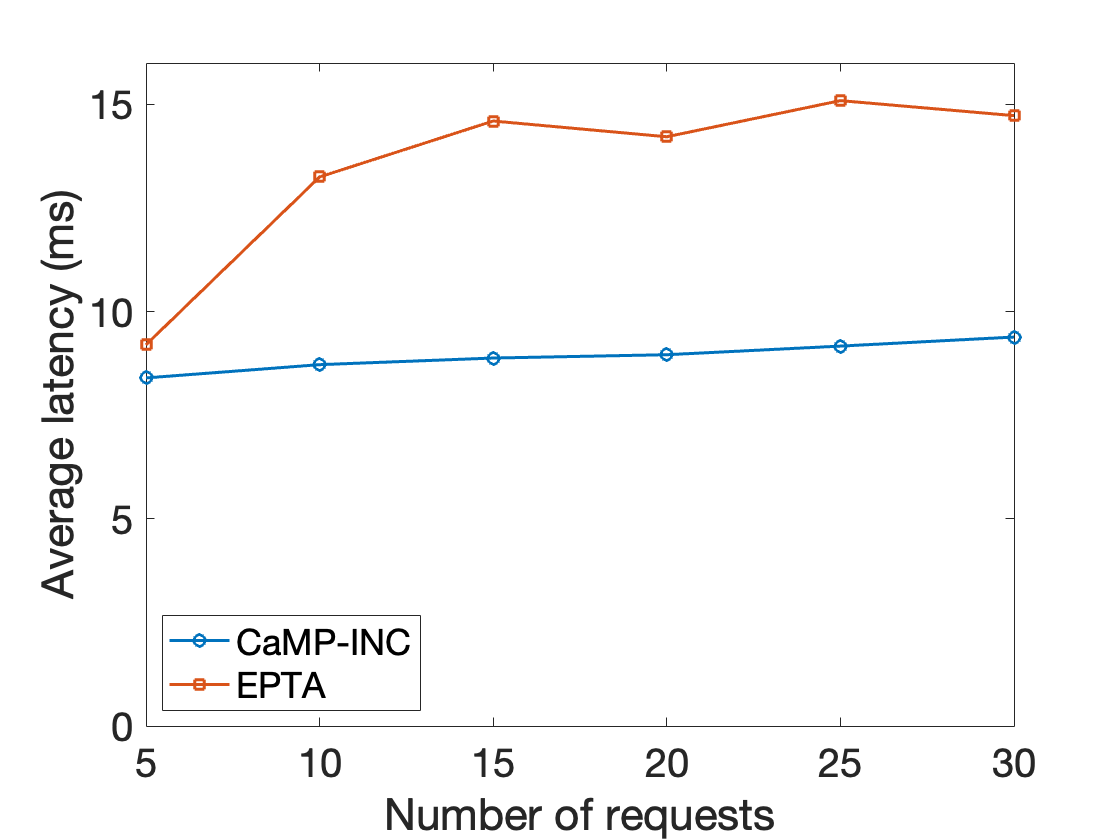}
    \caption{Average service request latency}
    \label{avg_latency}
  \end{subfigure}%
  \begin{subfigure}{.5\linewidth}
    \centering
    \includegraphics[scale=0.195]{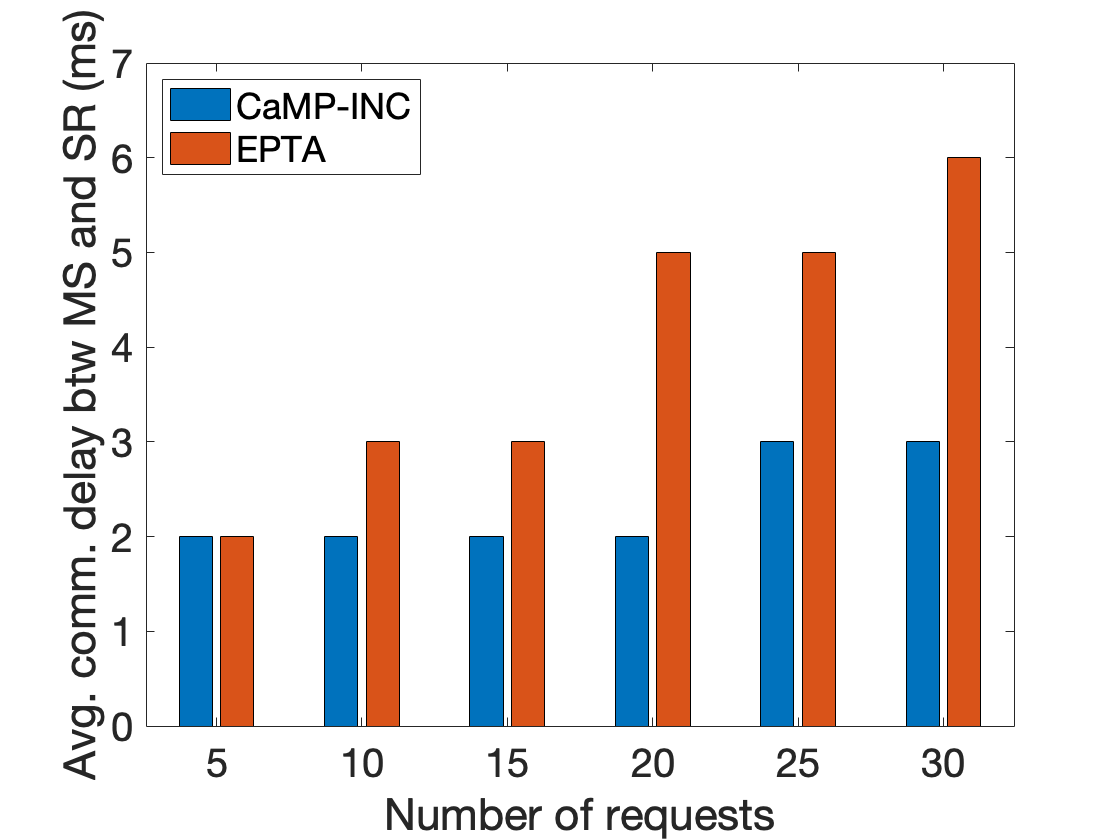}
    \caption{Average communication delay between microservices (MS) and service registries (SR)}
    \label{avg_distance}
  \end{subfigure}
  \caption{Impact of microservice architecture requirements}
\end{figure*}

\begin{figure}
\centering
\includegraphics[scale=0.195]{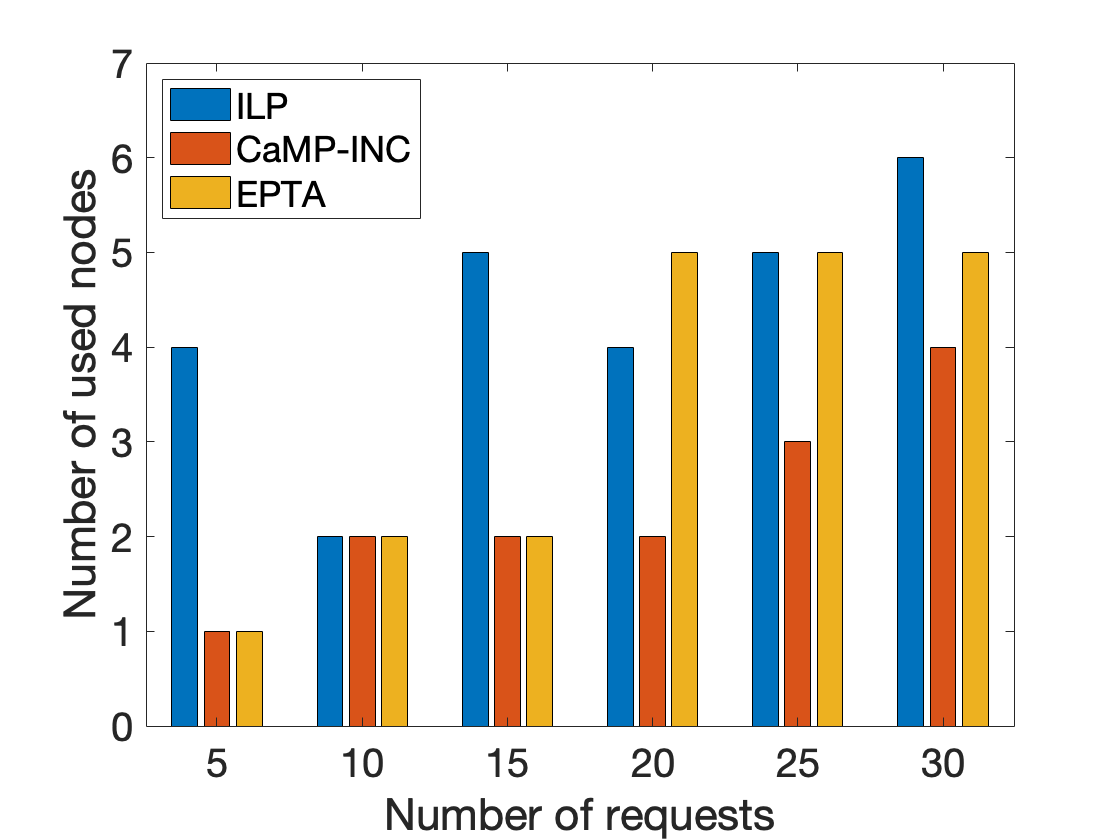}
    \caption{Number of nodes used for placement}
\label{used_nodes}
\end{figure}

Next, we compare the performance of CaMP-INC and EPTA regarding microservice architecture requirements. Fig.~\ref{avg_latency} depicts the impact of failure detection. We can observe that the average service request latency for CaMP-INC is slightly increasing with the increase of the number of requests since, the more requests there are, the more microservices are placed in longer paths and therefore link delays are added. 
Also, EPTA has a higher average service request latency than CaMP-INC because it does not consider failed nodes and subsequently may place microservices in failed nodes. When a request arrives at a failed node, the processing fails. Thus, the request is sent to the orchestrator to decide on another placement.
In order to evaluate the placement of the service registries, Fig.~\ref{avg_distance} shows the average communication delay between the placed microservices and the service registries. CaMP-INC places the service registries better than EPTA, which reduces the distance and provides lower communication delay between microservices and service registries. This is highly beneficial in delay constrained scenarios.

Fig.~\ref{used_nodes} shows the number of nodes used for placing the microservices. We observe that CaMP-INC allows to have fewer number of nodes for placement. It might be important when there is a need to keep some nodes for other processing tasks. However as expected, ILP uses the highest number of nodes to have the optimal cost.

\section{Conclusion}\label{conclusion}
Microservices adoption in the cloud and edge environments became very popular due to the attractive characteristics of microservice architecture. Similarly, microservices' lightweight and precise nature make them very convenient for performing in-network computing. In order to take more advantage of microservice architecture, it is mandatory to consider its requirements. This paper proposed a solution named CaMP-INC for Components-aware Microservices Placement for In-Network Computing Cloud-Edge Continuum. In this study, the problem is formulated as ILP, and a cost-efficient CaMP-INC algorithm is proposed. Compared to the ILP and the state-of-the-art EPTA solution, the results show that CaMP-INC reduces the execution time. Also, it achieves a good trade-off between execution time and total cost. We plan to extend our CaMP-INC solution by integrating the rest of microservice architecture requirements (i.e. microservices replication and load balancing) in future work.



\ifCLASSOPTIONcaptionsoff
  \newpage
\fi

\bibliographystyle{IEEEtran}
\bibliography{bare_jrnl}

\begin{thebibliography}{10}
\providecommand{\url}[1]{#1}
\csname url@samestyle\endcsname
\providecommand{\newblock}{\relax}
\providecommand{\bibinfo}[2]{#2}
\providecommand{\BIBentrySTDinterwordspacing}{\spaceskip=0pt\relax}
\providecommand{\BIBentryALTinterwordstretchfactor}{4}
\providecommand{\BIBentryALTinterwordspacing}{\spaceskip=\fontdimen2\font plus
\BIBentryALTinterwordstretchfactor\fontdimen3\font minus
  \fontdimen4\font\relax}
\providecommand{\BIBforeignlanguage}[2]{{%
\expandafter\ifx\csname l@#1\endcsname\relax
\typeout{** WARNING: IEEEtran.bst: No hyphenation pattern has been}%
\typeout{** loaded for the language `#1'. Using the pattern for}%
\typeout{** the default language instead.}%
\else
\language=\csname l@#1\endcsname
\fi
#2}}
\providecommand{\BIBdecl}{\relax}
\BIBdecl

\bibitem{KULATUNGA2017106}
C.~Kulatunga, K.~Bhargava \emph{et~al.}, ``Cooperative in-network computation
  in energy harvesting device clouds,'' \emph{Sustainable Computing:
  Informatics and Systems}, vol.~16, pp. 106--116, 2017.

\bibitem{Dan2019}
D.~R.~K. Ports and J.~Nelson, ``When should the network be the computers,'' in
  \emph{Proc. of Workshop on Hot Topics in Operating Systems(HotOS'19)}, 2019,
  pp. 209--215.

\bibitem{8585089}
M.~Mazzara, N.~Dragoni \emph{et~al.}, ``Microservices: Migration of a mission
  critical system,'' \emph{IEEE Trans. Serv. Comput.}, pp. 1--1, 2018.

\bibitem{hawilo2019exploring}
H.~Hawilo, M.~Jammal, and A.~Shami, ``Exploring microservices as the
  architecture of choice for network function virtualization platforms,''
  \emph{IEEE Network}, vol.~33, no.~2, pp. 202--210, 2019.

\bibitem{sampaio2019improving}
A.~R. Sampaio, J.~Rubin \emph{et~al.}, ``Improving microservice-based
  applications with runtime placement adaptation,'' \emph{J. Internet Serv.
  Appl.}, vol.~10, no.~1, pp. 1--30, 2019.

\bibitem{pallewatta2019microservices}
S.~Pallewatta, V.~Kostakos, and R.~Buyya, ``Microservices-based iot application
  placement within heterogeneous and resource constrained fog computing
  environments,'' in \emph{Proc. of IEEE/ACM Int. Conf. on Utility and Cloud
  Computing}, 2019, pp. 71--81.

\bibitem{messina2016database}
A.~Messina, R.~Rizzo \emph{et~al.}, ``The database-is-the-service pattern for
  microservice architectures,'' in \emph{Proc. Int. Conf. Information
  Technology in Bio-and Medical Informatics}.\hskip 1em plus 0.5em minus
  0.4em\relax Springer, 2016, pp. 223--233.

\bibitem{kakivaya2018service}
G.~Kakivaya, L.~Xun \emph{et~al.}, ``Service fabric: a distributed platform for
  building microservices in the cloud,'' in \emph{Proc. of EuroSys conf.},
  2018, pp. 1--15.

\bibitem{liu2019e3}
M.~Liu, S.~Peter \emph{et~al.}, ``E3:energy-efficient microservices on
  smartnic-accelerated servers,'' in \emph{Proc. USENIX Annual Technical
  Conf.}, 2019, pp. 363--378.

\bibitem{ding2022kubernetes}
Z.~Ding, S.~Wang, and C.~Jiang, ``Kubernetes-oriented microservice placement
  with dynamic resource allocation,'' \emph{IEEE Trans. Cloud Comput.}, no.~01,
  pp. 1--1, 2022.

\bibitem{kaur2022latency}
K.~Kaur, F.~Guillemin \emph{et~al.}, ``Latency and network aware placement for
  cloud-native 5g/6g services,'' in \emph{Proc. Consumer Commun. \& Netw. Conf.
  (CCNC)}.\hskip 1em plus 0.5em minus 0.4em\relax IEEE, 2022, pp. 114--119.

\bibitem{mpcsm}
Y.~Wang, C.~Zhao \emph{et~al.}, ``Mpcsm: Microservice placement for edge-cloud
  collaborative smart manufacturing,'' \emph{IEEE Trans. Industr. Inform.},
  vol.~17, no.~9, pp. 5898--5908, 2020.

\bibitem{cattrysse1992survey}
D.~G. Cattrysse and L.~N. Van~Wassenhove, ``A survey of algorithms for the
  generalized assignment problem,'' \emph{Eur. J. Oper. Res.}, vol.~60, no.~3,
  pp. 260--272, 1992.

\bibitem{page1999pagerank}
L.~Page, S.~Brin \emph{et~al.}, ``The pagerank citation ranking: Bringing order
  to the web.'' Stanford InfoLab, Tech. Rep., 1999.

\bibitem{EPTA}
X.~Wan, X.~Guan \emph{et~al.}, ``Application deployment using microservice and
  docker containers: Framework and optimization,'' \emph{J. Netw. Comput.
  Appl.}, vol. 119, pp. 97--109, 2018.

\end{thebibliography}





\end{document}